\documentclass[lettersize,journal]{IEEEtran}
\usepackage[numbers]{natbib}
\usepackage{lineno,hyperref}
\usepackage{graphicx}
\usepackage{epstopdf}
\usepackage{algorithm}
\usepackage{algorithmicx}
\usepackage{algpseudocode}
\usepackage{amsmath}
\usepackage{amstext}
\usepackage{amssymb}
\usepackage{caption}
\usepackage{subfigure}
\usepackage{color}
\usepackage{multirow}
\usepackage{subfigure}
\usepackage[english]{babel}%
\usepackage{tabularx}

\begin{document}

\title{A Unified Blockchain-Semantic Framework for Wireless Edge Intelligence Enabled Web 3.0 }

\author{Yijing Lin,
		Zhipeng Gao*,
		Hongyang Du,
		Dusit Niyato,
    		Jiawen Kang,
    		Ruilong Deng,
    Xuemin Sherman Shen
\thanks{Corresponding author: Zhipeng Gao (e-mail: gaozhipeng@bupt.edu.cn).}
\thanks{Yijing Lin and Zhipeng Gao are with the State Key Laboratory of Networking and Switching Technology, Beijing University of Posts and Telecommunications, China. Hongyang Du and Dusit Niyato are with the School of Computer Science and Engineering, Nanyang Technological University,
Singapore. Jiawen Kang is with the School of Automation, Guangdong University of Technology, China. Ruilong Deng is with the College of Control Science and Engineering, Zhejiang University, China. Xuemin Sherman Shen is with the Department of Electrical and Computer Engineering, University of Waterloo, Canada}


}

\maketitle

\begin{abstract}

Web 3.0 enables user-generated contents and user-selected authorities. With decentralized wireless edge computing architectures, Web 3.0 allows users to read, write, and own contents. A core technology that enables Web 3.0 goals is blockchain, which provides security services by recording content in a decentralized and transparent manner. However, the explosion of on-chain recorded contents and the fast-growing number of users cause increasingly unaffordable computing and storage resource consumption. A promising paradigm is to analyze the semantic information of contents that can convey precisely the desired meanings without consuming many resources. In this article, we  propose a unified blockchain-semantic ecosystems framework for wireless edge intelligence-enabled Web 3.0. Our framework consists of six key components to exchange semantic demands. We then introduce an Oracle-based proof of semantic mechanism to implement on-chain and off-chain interactions of Web 3.0 ecosystems on semantic verification algorithms while maintaining service security. An adaptive Deep Reinforcement Learning-based sharding mechanism on Oracle is designed to improve interaction efficiency, which can facilitate Web 3.0 ecosystems to deal with varied semantic demands. Finally, a case study is presented to show that the proposed framework can dynamically adjust Oracle settings according to varied semantic demands.

\end{abstract}


\section{Introduction}

\IEEEPARstart{T}{he} Web has become an inseparable part of human lives around the world since the unprecedented evolution of wireless communication technologies from 4G, 5G to 6G. There are three iterations of the Web, which are classified as Web 1.0, Web 2.0, and Web 3.0. Web 1.0 was created by Tim Berners-Lee in 1989 to construct information networks and provide users with static resources to read through centralized architectures. The current web, i.e., Web 2.0, is a term coined by Tim O'Reilly in 2007 \cite{o2007web}, which enables users to read and write dynamic contents through distributed architectures to form social networks. Recently, Web 3.0 is becoming a promising concept as the next generation of information infrastructures due to the development of cutting-edge technologies, e.g., blockchain, semantic communication, and wireless edge computing. Rather than implementing computation and data storage in centralized data centers, Web 3.0 leverages decentralized wireless edge computing architectures to offload compute and storage capacity to the wireless edge side close to users. More importantly, it permits users to read, write and own contents, e.g., text, image, and video, through decentralized wireless edge computing architectures to build a smarter and more socially, economically connected society.

Secure data storage and efficient information interaction have always been the focus of Web researches. To ensure the first goal, i.e., data security, Blockchain-based Web 3.0 was proposed by Ethereum co-founder Gavin Wood in 2014 \cite{wood_2014}, which makes extensive use of blockchain ecosystems to achieve user-generated contents and user-selected authorities. Specifically, unlike the current Web 2.0 where content is controlled by tech giants without users' involvement and permission, Web 3.0 uses blockchain to record on-chain content in a decentralized, transparent, and traceable manner. By now, both academia and industries have agreed that Web 3.0 should be enabled by blockchain to guarantee transparency, security, and efficiency. As the new generation of Internet applications continues to emerge, i.e., Metaverse, users can access Web 3.0 services through a virtual avatar provided by Metaverse. With the help of the integration of new technologies, Web 3.0 can secure data information without the intervention of third parties \cite{tang2022roadmap}. However, significant computing and storage resources are consumed to record the content. Information overload on Web 3.0 takes up too many resources on wireless edge devices and limits the number of devices covered by the Web 3.0 network.

The recent vigorous development of semantic communication has brought potential solutions to this problem and achieved the second goal of Web 3.0 systems, i.e., efficient information interaction. By analyzing the semantics of the content rather than the original raw data, the system can run more efficiently without affecting the completion of various service tasks. The data of wireless edge devices can be reduced to 2.5\% by semantic communication compared with traditional methods to enable efficient information exchange \cite{xie2020lite}. Users can also use and appreciate the semantic information and its services more directly and effectively. 

The analysis of content semantics in Web 3.0 can be traced back to 2001 when the World Wide Web inventor Tim Berners-Lee proposed Semantic Web 3.0~\cite{berners2001semantic}. Machine learning and artificial intelligence technologies are used to customize semantic demands. Specifically, the application of semantic mechanisms in Web 3.0 can bring the following benefits:
\begin{itemize}
    \item A proof of semantic mechanism can avoid user information overload and reduce network resource consumption.
    \item The content security can be further enhanced by verifying the semantics of the content.
    \item The efficiency of information interaction can be improved and the goal-oriented communication can be achieved, which reduces the latency and achieves the variety of Web 3.0 services.
\end{itemize}

Therefore, Web 3.0 is expected to convey the semantics of contents in decentralized wireless edge computing architectures, and connect blockchain and off-chain semantic ecosystems. However, even with the above benefits, the design of a blockchain-semantic ecosystem faces many difficulties. On the one hand, it is difficult for wireless edge devices to process and exchange extensive semantic information in the blockchain. The reason is that blockchain cannot invoke actively off-chain contents to achieve a consensus. Moreover, since on-chain resources are limited and expensive, semantic extraction algorithms cannot be executed by miners through smart contracts. These limitations cause insecure services on on-chain and off-chain interactions. On the other hand, a well-designed semantic verification mechanism can reduce additional network resource consumption in transmitting irrelative contents to consumers. The data exploration in the Web 3.0 era further challenges the performance of decentralized wireless edge computing architectures.

To solve the aforementioned problems, we are the first to propose a unified blockchain-semantic ecosystems framework for wireless edge intelligence-enabled Web 3.0. The framework contains six key components to consider the content semantics, and exploit Oracle to integrate blockchain and off-chain semantic ecosystems. Our contributions are summarized as follows:

\begin{itemize}
    \item To capitalize on the great benefits of semantic communication, we propose a unified blockchain and semantic ecosystems framework for wireless edge intelligence-enabled Web 3.0. To the best of our knowledge, this is the first work on wireless edge-intelligence enabled Web 3.0. We believe that this is a timely study, as the blockchain and semantic ecosystems are widely used in many scenarios, like Metaverse.
    \item To maintain service security, we design an Oracle-based proof of semantic mechanism to implement on-chain and off-chain interactions and transfer on-chain semantic verification algorithms to off-chain Oracle for reaching a consensus on semantic information in loosely trusted environments.
    \item To improve interaction efficiency, we establish an adaptive DRL-based sharding mechanism for verifiers of Oracle to cater varied semantic demands in dynamic Web 3.0 environments.
\end{itemize}

\section{Wireless Edge Intelligence-Enabled Web 3.0}

In this section, we first introduce technologies and applications related to Web 3.0. We then discuss how wireless edge intelligence enables Web 3.0. Finally, we demonstrate Web 3.0 research gaps as our motivation for designing the unified blockchain-semantic ecosystems framework.

\subsection{Introduction of Web 3.0}

From read-only Web 1.0 information networks and read-write Web 2.0 platform networks, Web 3.0 promises to be read-write-execute value networks \cite{beniiche2022society} based on blockchain, semantic ecosystems, and wireless edge computing, which can achieve user-generated content and user-selected authorities. Blockchain is widely considered to be a modern usage of Web 3.0 to provide a decentralized way to develop an open and transparent web experience for users. Besides, semantic ecosystems that can combine with other information infrastructures, like AI and 6G, provide a data-driven way to develop an intelligent and connected web experience for users. Moreover, wireless edge computing provides an efficient way to process data at the wireless network edge on devices for users, since intelligent devices permit the core network to decentralize computing and storage capabilities to the wireless edge to reduce the load. Current Web 3.0 mainly focuses on blockchain-enabled applications, including DIDs, DeFi, DAM, and Metaverse.

\textbf{Decentralized Identifiers (DIDs):} Web 3.0 enables data owners to maintain of the contents that they produced, and DIDs facilitate users of Web 3.0 to control information and privacy around themselves instead of storing entirely on centralized and propriety databases controlled by tech giants. DIDs are the cornerstone of other Web 3.0 technologies, like DeFi, DAM, and Metaverse, to exchange securely signed information in a user-centric and user-controlled way.

\textbf{Decentralized Finance (DeFi):} Web 3.0 supported by blockchain is used to create digital economies for traditional economies through tokenization, which reshapes the financial business. DeFi products and services are developed through smart contracts on blockchain to enable anyone more control to access and trade products and services through self-controlled wallets regardless of who or where they are. 

\textbf{Digital Asset Management (DAM)}: User-generated contents and authorities of Web 3.0 are digital assets, like cryptocurrency, non-fungible tokens, virtual real estate, and user avatars. DAM allows users to create, own, access, trade, and destroy their digital assets. In addition, governments can also benefit from DAM of Web 3.0 to reduce transaction costs in regulation.

\textbf{Metaverse:} User-centric perspectives of Web 3.0 facilitate users to immerse themselves in the world constructed by Metaverse and serve as the basis for connectivity of Metaverse, which is supported by blockchain, wireless edge computing, and interactive technologies.

\subsection{Wireless Edge Intelligence-Enabled Web 3.0}
To the best of our knowledge, relatively few work considered the integration of blockchain and semantic communication to implement wireless edge intelligence-enabled Web 3.0. The literature on Web 3.0 can be classified as blockchain Web 3.0 and semantic Web 3.0. A comparison between our paper and the related papers in the literature is shown in Table \ref{contributions}.

\begin{table*}[!t]
  \centering  
	\caption{A comparison contribution between relevant papers and our paper}  
	\label{contributions}  
\begin{tabular}{|c|c|p{13cm}|}
\hline
  \textbf{Type} & \textbf{Paper} & \textbf{Key Contributions of The Related Paper} \\
\hline
\hline
\multirow{3}{*}{\textbf{Blockchain Web 3.0}} &   \cite{beniiche2022society}        &    Introduce the impact of blockchain-based tokenomics on the interaction of humans with robot agents, discuss purpose-driven tokens and cyber-physical-social systems-based token engineering framework to motivate a paradigm of creating tokenized digital twins of assets for the future Web 3.0.                  \\
                                    &     \cite{liu2021make}       &             Define the era of Web 3.0 and introduce three key infrastructural enablers (individual blockchains, federated or centralized platforms, and interoperability platforms) for Web 3.0. The interoperability platform, HyperService, is proposed to build and execute Web 3.0 dApps across blockchains.                           \\ 
                                     &     \cite{aoun2021review}      &          Review Blockchain's impact on industries and economy, which eliminate most of functions that exist with Web 2.0 and reliy on third parties. Explore areas where blockchain technologies enable for Industry 4.0.                              \\
\hline
\multirow{3}{*}{\textbf{Semantic Web 3.0}}   
								&  \cite{atzori2020special}         &     An overview of discussing data exploration in the Web 3.0 age, including data mining, query languages, and semantic analysis. Introduce semantics and patterns of Web 3.0, which enables the Web readable by machines through constructing connections between new knowledge.                                 \\
                                    &   \cite{hiremath2016alteration}        &       Review the development and compare the characteristics of Web from Web 1.0, Web 2.0 and Web 3.0. Introduce Web 3.0 as an executable and semantic web to enable semantic-aware and context-aware multi-user virtual environments.                                \\
                                    &  \cite{ghelani2022conceptual}          &    Discuss the transformations of technologies and business models as with Web 3.0. Propose a conceptual framework for marketing with the development of Web 3.0. Explore necessary factors beyond technologies, including involvement and collaboration of human engagement.                                                        \\
\hline
\multicolumn{2}{|c|}{\textbf{How this paper differs}}  & These papers either focus on blockchain Web 3.0 or Semantic Web 3.0. Neither of them considers the integration of blockchain and semantic ecosystems. A unified blockchain-semantic ecosystems framework is important for Web 3.0, which can provide decentralization supported by blockchain and reduce user information overloading enabled by semantic ecosystems.  \\
\hline
\hline
\end{tabular}
\end{table*}

\textbf{Blockchain Web 3.0} is based on blockchain ecosystems to achieve Web 3.0 in a decentralized way. Researchers focus on different core layers of blockchain to facilitate Web 3.0. In the incentive layer, Abdeljalil Beniiche \textit{et al.} \cite{beniiche2022society} utilized blockchain to create tokenized digital twins of assets for the future token economy of Web 3.0. In the interoperability layer, Zhuotao Liu \textit{et al.} \cite{liu2021make} proposed HyperService, an interoperability platform, to deliver interoperability and programmability across heterogeneous blockchains for Web 3.0 applications to make Web 3.0 connected. Alain Aoun \textit{et al.} \cite{aoun2021review} proposed that blockchain technology is a foundation protocol for Web 3.0, which eliminates third parties supported by peer-to-peer networks. 

\textbf{Semantic Web 3.0} consists of semantic communication and other semantic-related technologies to construct Web 3.0 in an efficient way. Semantic communication enables producers to transmit relative semantic information to consumers to avoid information overloading. Researchers focus on two key components of the semantic web, including data and personalization, to support Web 3.0. Maurizio Atzori \textit{et al.} \cite{atzori2020special} considered data exploration in Web 3.0 Age and introduced human-computer interaction, semantics and patterns, recommender systems to demonstrate opportunities in the emerging Web 3.0 paradigm. B. K. Hiremath \textit{et al.} \cite{hiremath2016alteration} described the semantic web as a database serving user's experience and reshaping a new user-oriented, connected, and intelligent form. Diptiben \textit{et al.} \cite{ghelani2022conceptual} proposed semantic web allows Web 3.0 services to communicate with each other and implement individualized and behavioral forms. 

The current semantic communication is an end-to-end communication, while it is hard to connect users and share semantic information in loosely trusted environments. Besides, blockchain can facilitate information of semantic communication to release value. Therefore, it is necessary to connect blockchain and semantic ecosystems.

\subsection{Research Gaps of Web 3.0}
As mentioned in the above sections, many underlying technologies of Web 3.0 are still in their infancy. Moreover, the integrations of blockchain and semantic communication are overlooked by researchers from both academia and industry. The semantic-based Web 3.0 without blockchain cannot maintain decentralized deployments and ownership of contents. The blockchain-based Web 3.0 without semantic ecosystems cannot facilitate semantic contents to benefit from AI and 6G.

However, the integration of blockchain, semantic communication, and wireless edge computing still suffers from the following challenges: 1) Since semantic contents come from off-chain storage, blockchain cannot directly invoke or process semantic verification algorithms to verify semantic contents due to deterministic characteristics. Deterministic characteristics require miners to verify blockchain in a decentralized way to maintain safety. It is difficult for miners to execute off-chain unknown verification since different off-chain semantic information may cause inconsistency among miners. Therefore, how to maintain service security is the first challenge for Web 3.0 to implement on-chain and off-chain interactions on semantic information. 2) Semantic content should be shared with participants who have similar knowledge to release value. It is necessary to divide Web 3.0 users into multiple shards according to varied semantic demands. Static sharding strategies cannot adjust when semantic demands dynamically change. Therefore, how to improve interaction efficiency is the second challenge for Web 3.0 to adapt to dynamic semantic demands. To fill this research gap, we aim to design a unified blockchain-semantic ecosystems framework to support wireless edge intelligence-enabled Web 3.0 by designing a proof of semantic mechanism to facilitate semantic demands circulating in value networks and a dynamic sharding mechanism to adapt to varied semantic demands.

\section{Unified Blockchain-Semantic Ecosystems Framework for Wireless Edge Intelligence-Enabled Web 3.0}
\label{sec_framework}

\begin{figure*}[!t]
  \centering
  \includegraphics[width=7in]{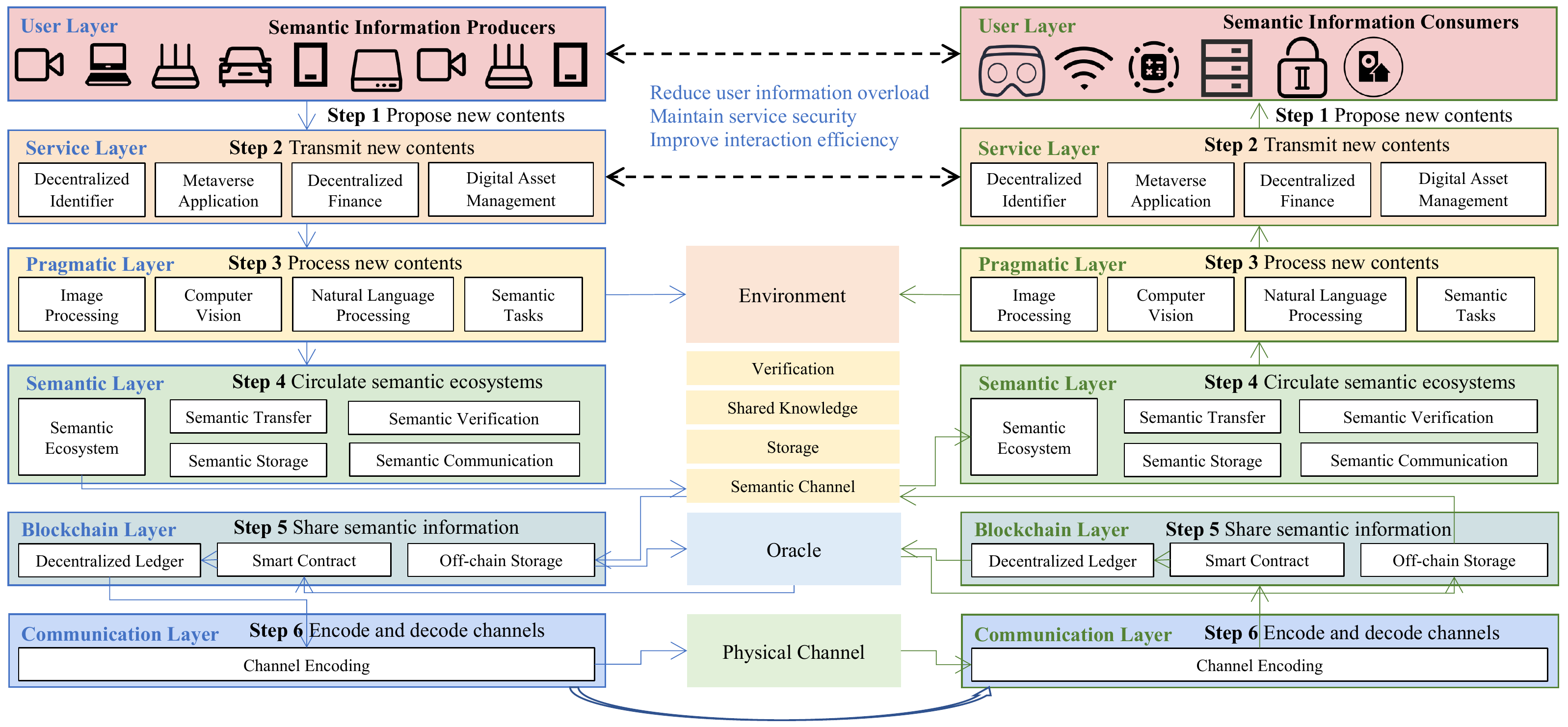}
  \caption{Illustrative Unified Blockchain-Semantic Ecosystems Framework for Wireless Edge Intelligence-Enabled Web 3.0}
  \label{fig_arch}
\end{figure*}

In this section, we first propose the architecture and workflows of the unified blockchain-semantic ecosystems framework for wireless edge intelligence-enabled Web 3.0 to reduce overloaded information. We introduce the proof of semantic mechanism to maintain service security and then design the adaptive DRL-based sharding mechanism to improve interaction efficiency given varied semantic demands.

\subsection{Architecture and Workflows}
As shown in Fig. \ref{fig_arch}, the architecture of the unified blockchain-semantic ecosystems framework for wireless edge intelligence-enabled Web 3.0 consists of six layers, including the user layer, service layer, pragmatic layer, semantic layer, blockchain layer, and communication layer. The user layer consists of wireless edge devices (semantic information producers and consumers) with limited resources, which cannot process complex semantic tasks. The service layer provides users with Web 3.0 services, such as DID, Metaverse, DeFi and NFT, where users can quickly use Web 3.0 services without being aware of underlying technologies. The pragmatic layer classifies tasks of semantic ecosystems for Web 3.0, such as image processing, computer vision, and natural language processing, which permits multiple types of semantic demands to encode and decode information. The semantic layer refers to semantic ecosystems, which is composed of semantic transfer, semantic verification, semantic storage, and semantic communication to compress and release semantic information efficiently. The blockchain layer enables Web 3.0 services in a decentralized, secure, and transparent manner, which is implemented by decentralized ledgers, smart contracts, off-chain storage, and Oracle to support on-chain and off-chain interactions. The communication layer realizes channel encoding and decoding through the physical channel to communicate semantic information producers with consumers. 

As shown in Fig. \ref{fig_arch}, the workflows of the proposed framework are described by the following steps.

\begin{itemize}
  \item \textbf{Step 1}: Propose new contents. Semantic information producers propose new contents through front-end interfaces of Web 3.0 services deployed on wireless edge devices to interact with consumers in the user layer. The edge devices include low-cost IoT or wearable devices, which have few resources to transmit and receive semantic contents in Web 3.0.   
  \item \textbf{Step 2}: Transmit new contents. Web 3.0 services receive contents through front-end interfaces and transmit them to the pragmatic layer for encoding and decoding semantic contents.  
  \item \textbf{Step 3}: Process new contents. Edge servers with strong computing and storage resources encode and decode contents according to different semantic demands and knowledge from environments, such as image processing, computer vision, and natural language processing.

  \item \textbf{Step 4}: Circulate semantic ecosystems. When each edge server receives contents, it first searches for local or shared knowledge to identify entity elements and logical relationships. When the contents cannot be recognized by local or shared knowledge, they start to train a new semantic model. To improve the search speed of semantic knowledge, each edge server should perceive environments related to communication before performing encoding and decoding semantic information. The encoded and decoded semantic contents can proceed with semantic transfer, semantic verification, semantic storage, and semantic communication operations when interacting with the blockchain layer through the semantic channel. The semantic transfer provides users with digital asset management to control ownership of contents. Semantic verification helps users to verify whether semantic contents precisely convey the desired meanings. Semantic storage permits users to offload computational-intensive semantic tasks to edge servers. Semantic communication permits users to exchange semantic content with shared knowledge for more efficient communications through the semantic channel. 

\begin{figure*}[!t]
  \centering
  \includegraphics[width=7in]{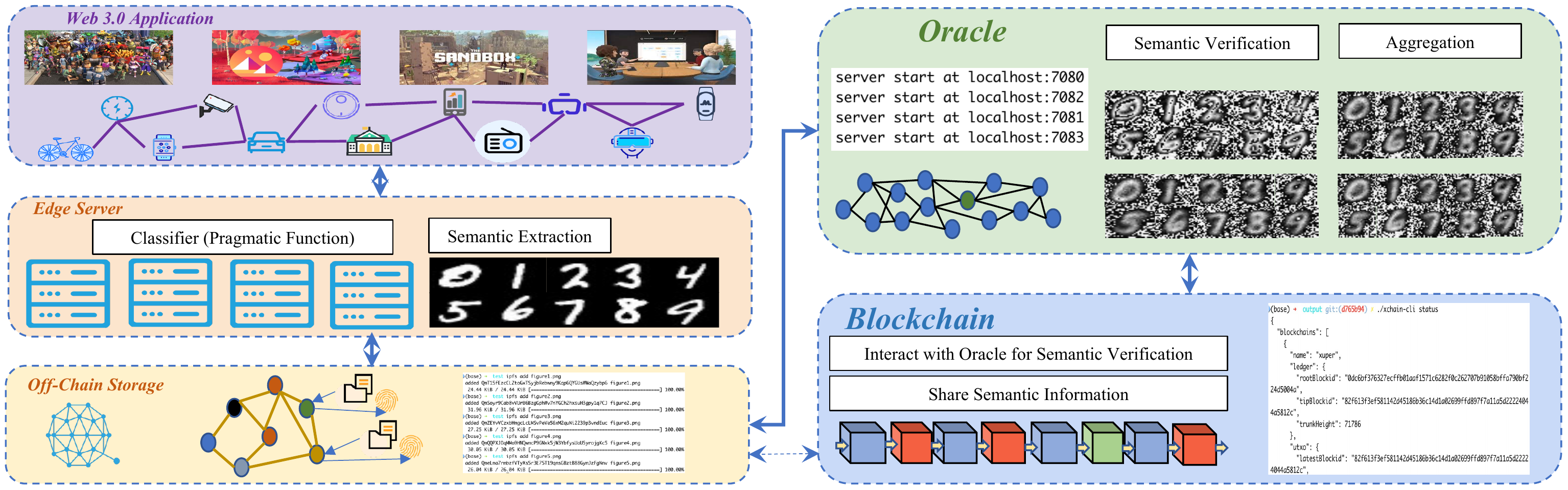}
  \caption{Illustrative Oracle-based Proof of Semantic Mechanism for Wireless Edge Intelligence-Enabled Web 3.0 }
  \label{fig_structure}
\end{figure*}

\begin{figure*}[!t]
  \centering
  \includegraphics[width=7in]{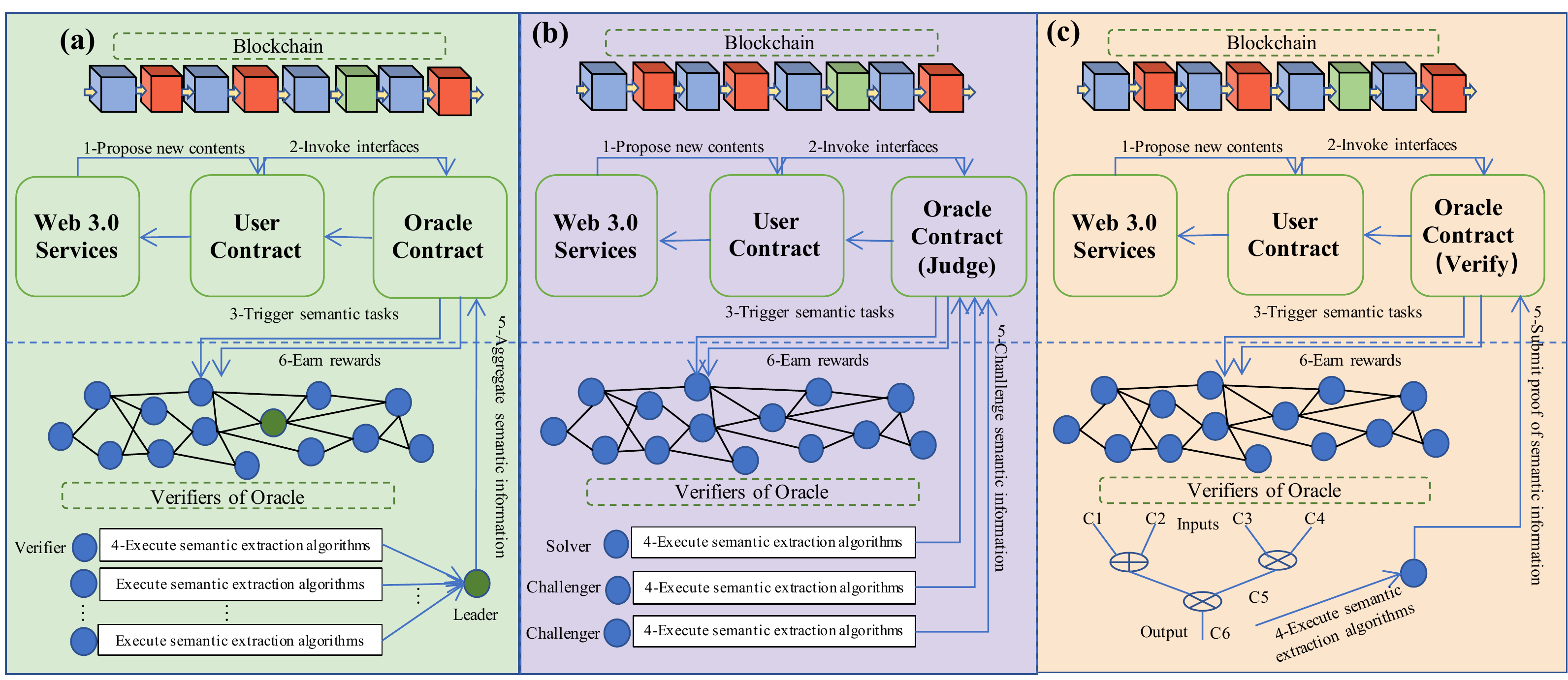}
  \caption{Illustrative Three Aggregation Mechanisms of Proof of Semantic}
  \label{fig_proof}
\end{figure*}

  \item \textbf{Step 5}: Share semantic information. Semantic information is recorded by blockchain through smart contracts deployed by Oracle in an active way when it is in off-chain storage provided by edge servers. Since blockchain cannot actively invoke off-chain semantic information, an Oracle-based proof of semantic mechanism is proposed to solve this issue in Section \ref{subsec_proof}. Moreover, since there are varied semantic demands, an adaptive DRL-based sharding mechanism on Oracle is proposed to adapt to this case in Section \ref{subsec_learning}
  \item \textbf{Step 6}: Encode and decode channels. Semantic information is transmitted by physical channels to interact with semantic information producers and consumers.
\end{itemize}


\subsection{Oracle-based Proof of Semantic Mechanism}
\label{subsec_proof}

Since edge servers are required to encode and decode semantic information in loosely trusted environments, blockchain is constructed among edge servers (i.e., miners) for semantic information sharing. The semantic information has to be verified before being added to a blockchain due to the blockchain's garbage-in garbage-out challenge. However, the semantic verification algorithm implemented in edge servers will result in inconsistent results due to different background knowledge, which is difficult for miners to verify semantic information and reach a consensus. Besides, since blockchain requires all miners to repeat smart contracts to verify correctness, it is impossible to execute complex on-chain semantic verification algorithms. Moreover, semantic verification algorithms require shared knowledge in off-chain storage to verify semantic information, while blockchain cannot actively invoke off-chain knowledge to proceed with the process. Therefore, we transfer on-chain semantic verification to off-chain Oracle to aggregate and verify semantic information from different knowledge and report aggregated semantic information to the blockchain to reach a consensus, as shown in Fig. \ref{fig_structure}. 

\begin{figure*}[!t]
  \centering
  \includegraphics[width=7in]{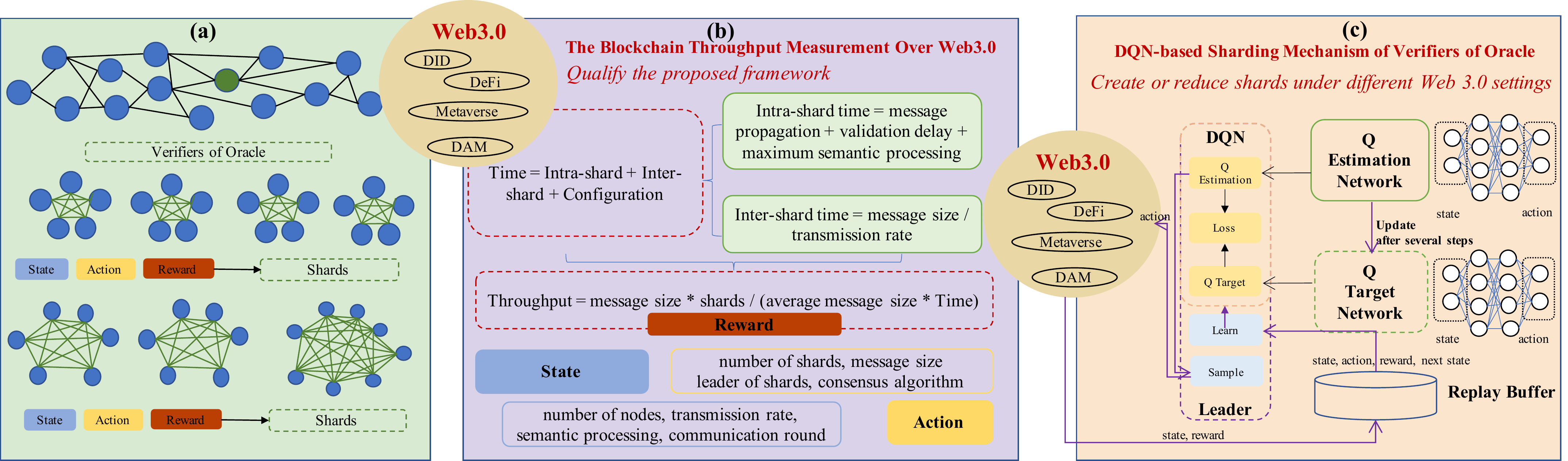}
  \caption{Illustrative Adaptive DRL-based Sharding Mechanism for Wireless Edge Intelligence-Enabled Web 3.0}
  \label{fig_learning}
\end{figure*}

Oracle \cite{breidenbach2021chainlink} is also a decentralized network composed of edge servers (verifiers). Oracle is required to deploy smart contracts (Oracle Contract) in advance for verifiers to subscribe to semantic tasks. New contents can be written into smart contracts (User Contract) and invoke interfaces provided by Oracle Contract to proceed with semantic tasks. Verifiers listen to events triggered by Oracle Contract and execute semantic verification algorithms to verify semantic information. Since there are multiple verifiers with different background knowledge, it is hard for the verifiers to obtain the same result about semantic verification. Therefore, there are off-chain and on-chain aggregation mechanisms to maintain the accuracy of semantic information, as shown in Fig. \ref{fig_proof}. 

For the off-chain aggregation mechanism, the verifiers select a leader to aggregate the results of semantic verification. The selection of leaders can refer to consensus algorithms. The verifiers are required to broadcast semantic verification results to the leader, while the leader utilized shared knowledge to verify the accuracy of semantic verification results. If the accuracy exceeds a predefined threshold, the semantic verification result provided by the verifiers will be aggregated by the leader. Once the leader submits the aggregated semantic verification results to the blockchain for consensus, the verifiers whose semantic verification results are aggregated will earn rewards (e.g., tokens paid by semantic information producers through smart contracts), as shown in Fig. \ref{fig_proof} (a).

For on-chain aggregation mechanisms, there is interactive or non-interactive verifications. Interactive verification refers to challengers executing off-chain semantic verification algorithms to challenge semantic verification results provided by solvers. Once the challengers succeed to reverse or equal the semantic verification results of solvers, challengers earn rewards, as shown in Fig. \ref{fig_proof} (b). Non-interactive verification generates arithmetic circuits of semantic verification algorithms and submits proof of semantic verification results based on the zero-knowledge proof to the blockchain for verification. Smart contracts verify proofs and distribute rewards, as shown in Fig. \ref{fig_proof} (c).

\subsection{Adaptive DRL-based Sharding Mechanism on Oracle}
\label{subsec_learning}

The primary precondition for semantic ecosystems is that all participants (including semantic information producers and consumers) have the same or similar background knowledge, including semantic entity elements and logical relationships. To connect blockchain and semantic ecosystems in a flexible way, the following challenges have to be solved.

\begin{itemize}
	\item How can verifiers of Oracle in the blockchain layer adapt to varied semantic demands?
	\item How can static sharding strategies with predetermined fixed settings to meet dynamic semantic demands if constructing multiple shards for verifiers of Oracle?
\end{itemize}

Therefore, we propose a DRL-based sharding mechanism for verifiers of Oracle in the blockchain layer to create or reduce shards under different Web 3.0 settings, which can dynamically adapt to different types of semantic tasks and improve the transaction processing speed of verifiers in Oracle, as shown in Fig. \ref{fig_learning}.

Since verifiers maintain decentralized deployments in Web 3.0, they cannot be scheduled by dynamic sharding strategies to reach a consensus. Therefore, the selection of dynamic sharding strategies is realized by the leader of verifiers of Oracle in the blockchain layer. When the leader implements the above process to obtain an optimal sharding setting, it can package the optimal setting into a message, and broadcast the message to other verifiers for a consensus. Other verifiers verify the message and extract the optimal setting to reach a consensus on the sharding strategies, as shown in Fig. \ref{fig_learning} (a).

The dynamic strategy regarding varied semantic demands is utilized to improve the throughput (the transaction processing speed) of verifiers of Oracle in the blockchain layer. The throughput is affected by the message size which is propagated in networks, the number of shards, average message size, and message process time. The message processing time includes intra-shard time, inter-shard time, and configuration time. The configuration time involves the shard formation phase, where the leader implements DRL processes to obtain optimal shard settings and reach a consensus on other verifiers. The intra-shard time comprises message propagation, validation delay, and the maximum semantic processing time to complete semantic tasks. The inter-shard time is calculated by the ratio of message size and transmission rate when verifiers of Oracle submit off-chain aggregation results to the blockchain to earn rewards, as shown in Fig. \ref{fig_learning} (b).

Considering the dynamic conditions of Web 3.0, we use a Deep Q-learning Network (DQN)-based sharding mechanism for verifiers of Oracle to improve the transaction speed. The action considers the number of nodes, message transmission rate, and semantic processing time and communication round to adapt to dynamic Web 3.0 environments. The state considers the number of shards, message size, leader of shards, and consensus algorithm to react to actions. The transaction speed is utilized as a reward given different states and actions. 

DQN introduces Q estimation and Q target networks with the same structure. The leader interacts with Web 3.0 environments to obtain the state transition (state, action, reward, next state) and store it in the replay buffer. When there are sufficient state transitions in the replay buffer, the leader randomly samples a minibatch of transitions from the replay buffer to calculate the rewards of the Q estimation and Q target networks. Comparing the difference in rewards, parameters of the Q estimation network are updated. After several steps, parameters of the Q target network are updated to obtain optimal sharding settings, as shown in Fig. \ref{fig_learning} (c).

\section{Case Study}

To evaluate the performance of the proposed unified blockchain-semantic ecosystems framework for wireless edge intelligence-enabled Web 3.0, we implement the simulation case based on a well-known gym framework \cite{openai} by using Pytorch to illustrate rewards under different Web 3.0 conditions, as exhibited in Fig. \ref{fig_learning}. In the constructed simulation network, the latency for configuring shards, the message validation delay, the average message size, the maximum semantic processing time, and the minimum transmission rate are 0.001s, 0.1s, 1MB, 20s, and 10Mbps, respectively. We utilize PBFT \cite{castro1999practical} as an example to explain the message propagation time = 2$*$(number of nodes in a shard)$*$(number of nodes in a shard $-$ 1)$*$message size $/$ transmission rate. The policy network of the leader adopts a two-layer fully connected network and ReLU activation function. The learning rate, the discount factor, the exploration rate, the batch size, the step interval, and the training epochs are 0.002, 0.98, 0.1, 64, 10, and 1000, respectively.

\begin{figure}[!t]
  \centering
  \subfigure[]{\includegraphics[width=1.5in]{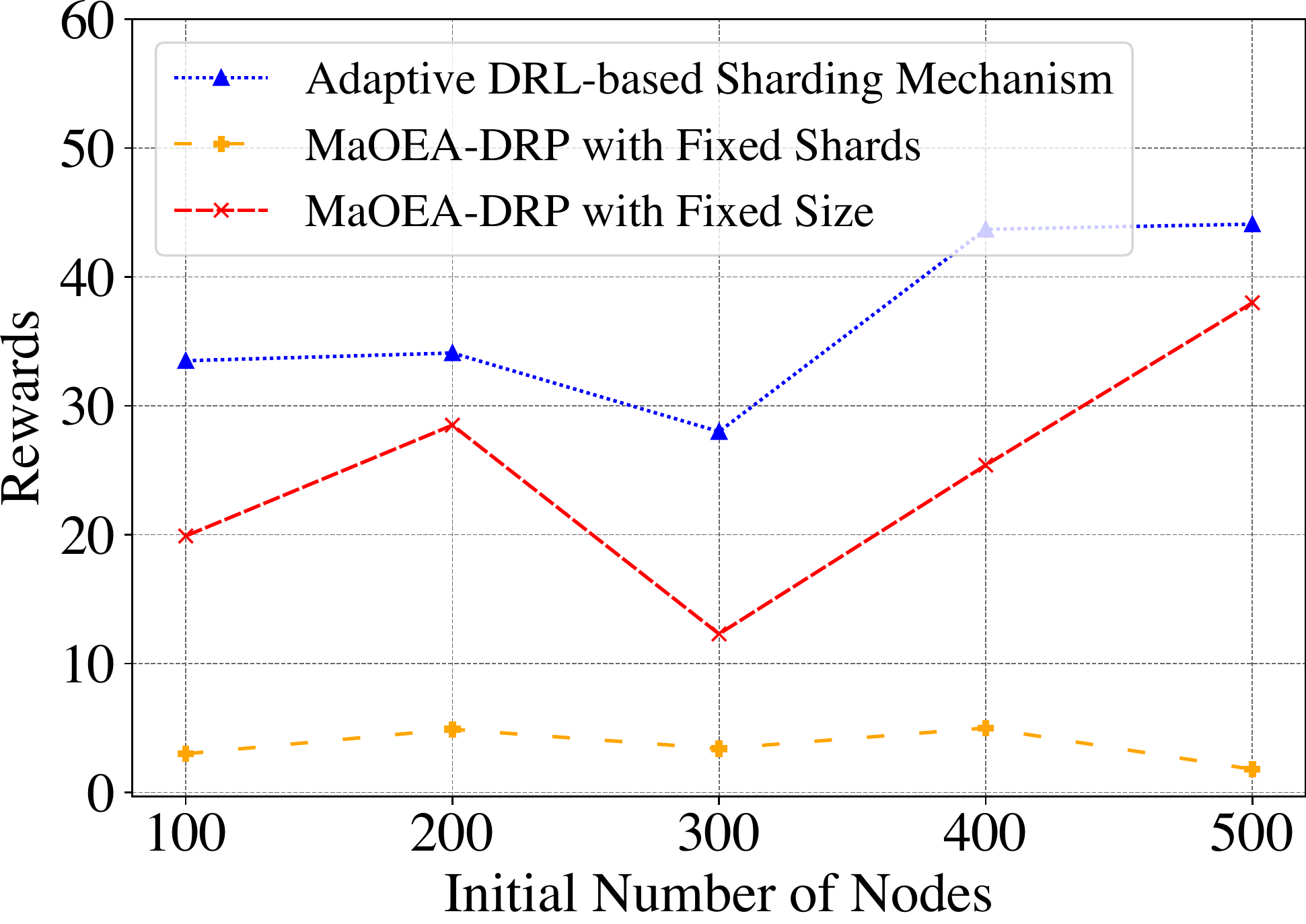}}
  \subfigure[]{\includegraphics[width=1.5in]{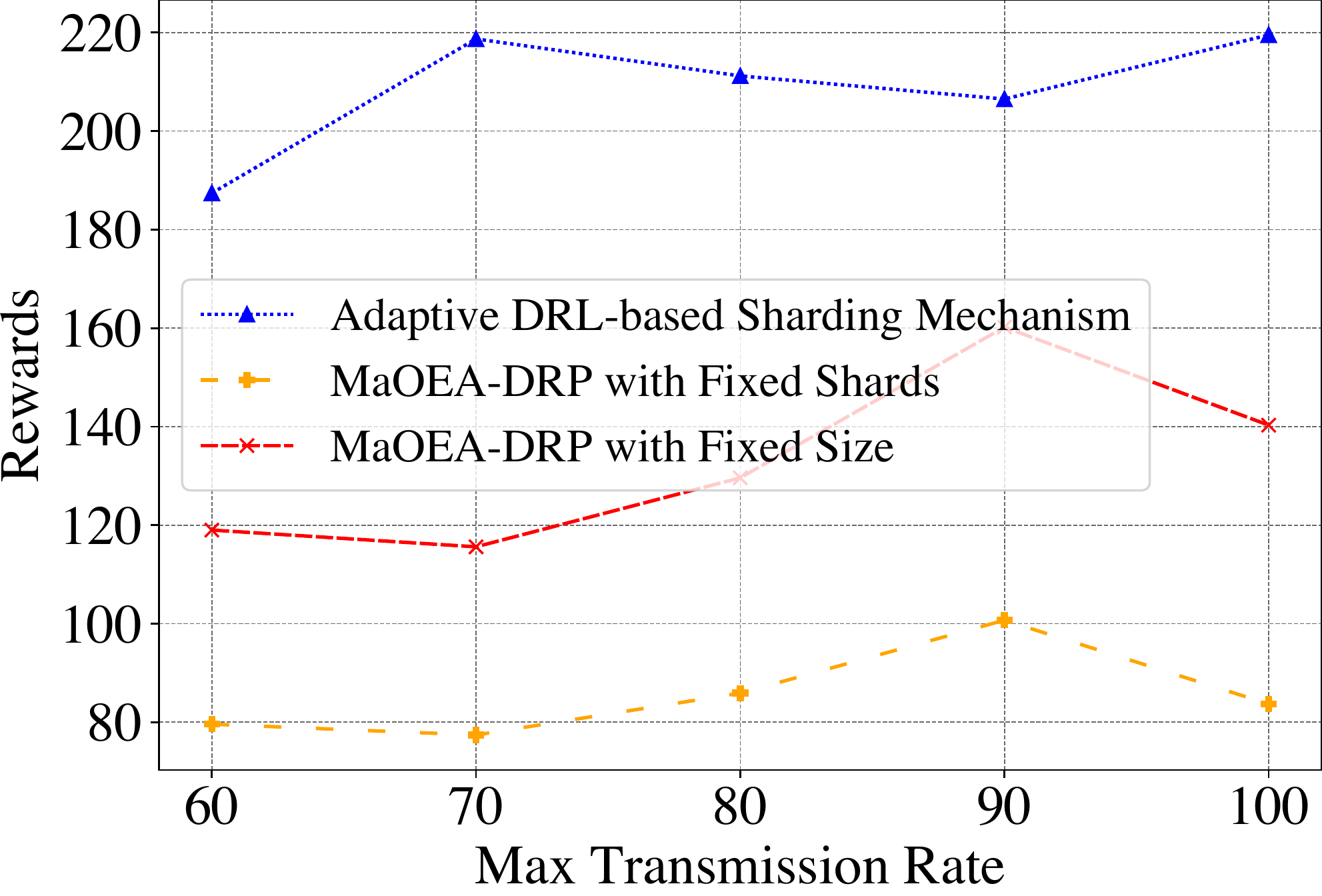}}
  \caption{An Illustration of Rewards under Different Web 3.0 Conditions}
  \label{fig_exp}
\end{figure}

Figure \ref{fig_exp} shows the performance of rewards for the proposed framework under dynamic Web 3.0 settings, including the inital number of nodes [100,500,100] and the maximum transmission rate [60,100,10]. The proposed framework is compared with the schemes that are currently in use, i.e., MaOEA-DRP \cite{cai2021sharding} with fixed shards (maximum number of shards) and message size (maximum message size). We can see that the proposed framework performs better than MaOEA-DRP under different Web 3.0 settings. This is because the proposed framework can increase or decrease adaptively number of shards and message size to deal with different initial number of nodes and max transmission rate, which can improve the rewards of the proposed framework.

\section{Future Research Directions}


\subsection{Web 3.0 Security and Privacy}

As noted in Section \ref{sec_framework}, blockchain and semantic ecosystems play important roles in the proposed framework. Security and privacy issues with these components can lead to serious consequences. For example, DeFi lost at least \$10 billion to hacks and fraud in 2021. Moreover, attackers can inject poisonous knowledge into semantic ecosystems to drag down the framework. Therefore, it is necessary to audit underlying codes of smart contracts and utilize federated learning or 
zero-knowledger proof to protect security and privacy before operating Web 3.0 services.

\subsection{Web 3.0 Management and Integration}

Since the semantic layer requires local and shared knowledge from edge servers to encode and decode semantic information, it is necessary to manage heterogeneous knowledge specifications for wireless edge intelligence-enabled Web 3.0 services. The proposed framework utilizes dynamic sharding mechanisms to classify the same semantic demands to solve the above issue, while it also requires some mathematical models for semantic information to unify knowledge specifications. Besides, the proposed framework requires the integration of new technologies, like 6G and quantum computing, to significantly improve communication and computation capabilities.

\subsection{Web 3.0 Scalability and Interoperability}

The throughput of blockchain is limited by consensus algorithms. Moreover, there are at least 1,000 blockchains networks with heterogeneous architectures. Therefore, the future Web 3.0 framework should construct more efficient consensus algorithms irrelative to network volumes while maintaining decentralization, and integrating cross-chain technologies into the unified blockchain-semantic ecosystems framework to facilitate Web 3.0 services to circulate around multiple blockchains.

\subsection{Web 3.0 Authentication and Governance}

Web 3.0 permits users to read, write and own content. It is necessary to facilitate data exchanges to release data value. Therefore, data governance should be permitted in the blockchain and semantic ecosystems. However, it is difficult to verify data authentication in Web 3.0 since counterparts of transactions are decentralized. Besides, data is easy to be copied in Web 3.0. Therefore, Secure Multi-Party Computation is necessary for Web 3.0 to implement data governance in blockchain and semantic ecosystems.

\section{Conclusion}

In this article, we have proposed an integrated framework that connect blockchain and semantic ecosystems for wireless edge-intelligence enabled Web 3.0 services, which avoids information overloading to human users. Under the proposed framework, an Oracle-based proof of semantic mechanism is introduced to transfer on-chain computation to off-chain Oracle for maintaining service security. Besides, we have designed an adaptive DRL-based sharding mechanism on Oracle to improve interaction efficiency. Moreover, we have illustrated a case to display the effectiveness of the proposed framework via simulation results. Finally, we have discussed the remained challenges and proposed potential solutions to address these issues.

\bibliographystyle{IEEEtran}
\bibliography{ref}









\end{document}